# Magnetic properties of (Fe$_{1-x}$Mn$_x$)$_2$AlB$_2$ and the impact of substitution on the magnetocaloric effect

D. Potashnikov,[1,2] E.N. Caspi, [3,4,5] A. Pesach, [3] S. Kota,[4] M. Sokol,[4,6] L.A. Hanner,[4] M.W. Barsoum,[4] H.A. Evans,[5] A. Eyal,[1] A. Keren[1] and O. Rivin[3]

August 2020

[1] *Faculty of Physics, Technion - Israeli Institute of Technology, Haifa 32000, Israel*

[2] *Israel Atomic Energy Commission, P.O. Box 7061, Tel-Aviv 61070, Israel*

[3] *Department of Physics, Nuclear Research Centre-Negev, P.O. Box 9001, Beer Sheva 84190, Israel*

[4] *Department of Materials Science and Engineering, Drexel University, Philadelphia, Pennsylvania 19104, USA*

[5] *Center for Neutron Research, National Institute of Standards and Technology, Gaithersburg, Maryland 20899, USA.*

[6] *Department of Materials Science and Engineering, Tel Aviv University, Ramat Aviv 6997801, Israel*

## Abstract

In this work, we investigate the magnetic structures of (Fe$_{1-x}$Mn$_x$)$_2$AlB$_2$ solid-solution quaternaries in the $x = 0$ to 1 range using x-ray and neutron diffraction, magnetization measurements, and mean-field theory calculations. While Fe$_2$AlB$_2$ and Mn$_2$AlB$_2$ are known to be ferromagnetic (FM) and antiferromagnetic (AFM), respectively, herein we focused on the magnetic structure of their solid solutions, which is not well understood. The FM ground state of Fe$_2$AlB$_2$ becomes a canted AFM at $x \approx 0.2$, with a monotonically diminishing FM component until $x \approx 0.5$. The FM transition temperature ($T_C$) decreases linearly with increasing $x$. These changes in magnetic moments and structures are reflected in anomalous expansions of the lattice parameters, indicating a magnetoelastic coupling. Lastly, the magnetocaloric properties of the solid solutions were explored. For $x = 0.2$ the isothermal entropy change is smaller by 30% than it is for Fe$_2$AlB$_2$, while the relative cooling power is larger by 6%, due to broadening of the temperature range of the transition.

## I. Introduction

The discovery of a giant magnetocaloric effect (MCE) near room temperature (RT) in Gd$_5$(Si$_2$Ge$_2$)[1] sparked an increasing interest in magnetic-based refrigeration. The numerous advantages of magnetic-based refrigeration include the elimination of moving parts and harmful gases. This method, known as active magnetic regeneration, is thus more efficient and environmentally "greener" compared to the current gas compression technology.[2] However, most of the known materials exhibiting a giant MCE near RT contain Gd, or other rare earths, that are too expensive for mass production. Therefore, research in the field has gravitated to magnetic materials containing more abundant elements such as the transition metals (TM) with magnetic ordering temperatures near RT. Examples include FeMnP$_{1-x}$As$_x$,[3] and Mn$_{1.25}$Fe$_{0.70}$P$_{1-x}$S$_x$,[4] which have tunable magnetic ordering temperatures and high magnetic entropy changes, and Ni-Mn-Sn Heusler alloys, which show a giant inverse MCE.[5]

Recently, the TM borides with the chemical formula $M_2$AlB$_2$, where $M$ = (Fe, Mn, Cr) have attracted much interest.[6,7] The compounds in this family (also called MAB phases) crystallize in the orthorhombic *Cmmm* space group with slabs of $M_2$B$_2$ stacked in between Al layers along the $b$ axis. Magnetic studies on the MAB phases have revealed that Fe$_2$AlB$_2$ orders ferromagnetically (FM) below ≈300 K,[8] Mn$_2$AlB$_2$ orders antiferromagnetically (AFM) below ≈313 K,[9] and Cr$_2$AlB$_2$ is paramagnetic (PM).[10] The near-RT FM phase transition of Fe$_2$AlB$_2$, along with it being composed of entirely earth-abundant and nontoxic elements, renders it a potential candidate for magnetic refrigeration. A large number of MCE studies in Fe$_2$AlB$_2$ are available (see Ref. 7 and references therein), which measure an isothermal entropy change of ≈4 J/kg K and an adiabatic temperature change of ≈2 K due to an applied field ($H$) of 2 T.

In an attempt to improve the available MC properties, several studies of MAB solid solutions, on both the $M$ and/or $A$ sites, were carried out.[11–16] For example, studies on the solid solution (Fe$_{1-x}$Mn$_x$)$_2$AlB$_2$ have shown that the addition of Mn gradually decreases the FM moments and the FM transition temperature. At intermediate Mn concentrations, the magnetic structure is hypothesized to be either a spin glass[12] or a disordered ferrimagnet,[13] due to competing magnetic interactions, but it has yet to be directly observed. The addition of AFM interactions is also known to widen the temperature range of the magnetic transition,[17] and thus allow for additional control over the MCE in the solid solution.

In order to further understand the magnetic properties of the (Fe$_{1-x}$Mn$_x$)$_2$AlB$_2$ system and enable fine tuning of its magnetic properties, we investigated the magnetic phase diagram of this system using x-ray, neutron diffraction and magnetization measurements. The measurements are qualitatively explained by a mean-field calculation of the magnetic phase diagram in the $x$–$T$ plane.

## II. Experimental Details

### A. Sample preparation and characterization

All compositions were prepared via a two-step reactive powder metallurgy route in a horizontal alumina tube furnace under flowing argon, Ar, as described in detail in the Supplemental Material (SM).[18] Samples with $^{11}$B (Cambridge Isotopes, 98%) were made with nominal Mn concentrations of $x$ = 0, 0.05, 0.1, 0.2, 0.25, 0.5, and 1. Additionally, samples with natural B were made with nominal Mn concentrations of $x$ = 0.2, 0.3, 0.5, 0.75, and 1.

X-ray powder diffraction (XRD) of (Fe$_{1-x}$Mn$_x$)$_2$Al$^{11}$B$_2$ ($x$ = 0, 0.05, 0.1, 0.2, 0.25, 0.5, 1) was performed using a Rigaku SmartLab diffractometer, equipped with a Cu K$\alpha$ radiation source and detector-side graphite monochromator. Additional samples with natural B ($x$ = 0.5, 0.75 and 1) were also measured. A step size of 0.015° and 6–8 s of dwell time per step was used in all cases.

The samples with $x = 0.5$ and 0.75 were also measured using the low background Bruker D8-Advance diffractometer, using CuKα radiation and an angular range of 10°–100° in steps of 0.01°.

## B. Magnetic properties

Magnetization measurements were performed using a Quantum Design MPMS3 system at the Quantum Material Research center in the Technion. Zero-field-cooled (ZFC) and field-cooled (FC) temperature scans were performed under a magnetic field ($H$) of 50 Oe ($\mu_0$ Oe = $10^{-4}$ Tesla). Field scans at constant temperatures were also carried out using $H$ in the 0–70 kOe range.

## C. Neutron powder diffraction

Three of the samples containing the $^{11}$B powders with $x = 0$, 0.1, and 0.2 were measured in the temperature ranges of 8–350 K ($x = 0$) and 8–300 K ($x = 0.1, 0.2$) using the BT-1 diffractometer at the National Center for Neutron Research located at the National Institute for Standards and Technology, USA. An incident wavelength of 2.079 Å was obtained using the Ge(311) monochromator and an in-pile collimation of 60'. The samples were loaded into a vanadium holder with a diameter of 9.2 mm. Two additional powders with $x = 0.25$ and 0.5, were measured using the KANDI-II diffractometer at the Israel Research Reactor II located at the Nuclear Research Center Negev, Israel.[19] The $x = 0.5$ sample was measured at 3, 100, 200, and 298 K, while the $x = 0.25$ sample was measured at 3 and 298 K.

## III. Theory

The magnetic properties of the $(Fe_{1-x}Mn_x)_2AlB_2$ system were modelled in the framework of the mean field theory (MFT) as described in Ref. 20. As noted above, the $M_2AlB_2$ unit cell has the orthorhombic *Cmmm* symmetry, where the $M$, Al and B atoms occupy the 4*j*, 2*a*, and 4*i* sites, respectively [Fig. 1(a)]. The magnetic ground states of the end compounds were previously determined by neutron diffraction to be FM for $Fe_2AlB_2$ [Fig. 1(b)] and AFM for $Mn_2AlB_2$ [Fig. 1(c)].[8,21] In the former case the magnetic moments are oriented along the crystallographic *a* axis. In the latter case, the magnetic unit cell is twice the size of the chemical unit cell along the *c* axis [propagation vector **k** = (0, 0, 1/2)].[21] The four magnetic moments in the chemical unit cell are all parallel and point along the crystallographic *b* axis.[9]

The reported possibility for low-dimensional magnetism[9] and canting of the Mn moments[21] in $Mn_2AlB_2$ was not taken into account in the present study. As discussed in the next sections, the estimated canted FM moment of ~8 ×$10^{-3}$ $\mu_B$[21] is two orders of magnitude below the detection limit of the neutron powder diffraction (NPD), and therefore cannot be observed by this method. $Mn_2AlB_2$ shall therefore be treated as a simple AFM. Since the four moments in the chemical unit cell are all parallel and equivalent in the magnetic ground states of the end compounds, a simplified description of the $M_2AlB_2$ compounds is obtained by averaging the four magnetic moments in the chemical unit cell into a single super moment. This simplifies the magnetic sublattice into a primitive orthorhombic Bravais lattice. Furthermore, in the zeroth approximation of the MFT,[22,23] the nearest neighbors along different crystallographic axes cannot be distinguished. The relative magnitudes of the exchange constants $J_{ij}$ along the $a$, $b$, and $c$ crystal axes are then taken to be equal [cubic lattice approximation, Fig. 1(d)].

To allow the description of an AFM unit cell, the cubic lattice is split into two sublattices ($A$ and $B$) along the $c$ axis [Fig. 1(d)]. The crystal directions are denoted using $\alpha$ and $\beta$. The single site Hamiltonian is then given by

$$\begin{aligned}\hat{H}_{\text{Fe},A} = -&\left\{\sum_{c_{AA}} z^{(c_{AA})}\left[(1-x)J_{\text{Fe-Fe}}^{(c_{AA}),\alpha\beta}\langle\hat{S}_{\text{Fe},A}^{\beta}\rangle + xJ_{\text{Fe-Mn}}^{(c_{AA}),\alpha\beta}\langle\hat{S}_{\text{Mn},A}^{\beta}\rangle\right]\right.\\ &\left.+\sum_{c_{AB}} z^{(c_{AB})}\left[(1-x)J_{\text{Fe-Fe}}^{(c_{AB}),\alpha\beta}\langle\hat{S}_{\text{Fe},B}^{\beta}\rangle + xJ_{\text{Fe-Mn}}^{(c_{AB}),\alpha\beta}\langle\hat{S}_{\text{Mn},B}^{\beta}\rangle\right]+\mu_{B}g_{\text{Fe}}^{\alpha\beta}H^{\beta}\right\}\hat{S}_{\text{Fe},A}^{\alpha}\\ =&-A_{\text{Fe}}^{\alpha}\hat{S}_{\text{Fe},A}^{\alpha},\end{aligned} \quad (1)$$

where $c_{AA}$ denotes the intrasublattice coordination shells, $c_{AB}$ denotes the intersublattice coordination shells, $z^{(c)}$ is the coordination number of the $c$th shell, $x$ is the Mn occupancy, $J_{M\text{-}M'}^{(c),\alpha\beta}$ are the anisotropic exchange constants of the $c$th coordination shell between $M$ and $M'$ atom types, $\mu_B$ is the Bohr magneton, $g_M^{\alpha\beta}$ is the anisotropic magnetic $g$-factor of atom type $M$, $H$ – is the applied field, $\hat{S}_{M,A}^{\alpha}$ is the spin operator of atom type $M$, on sub-lattice $A$, and $\langle\ \rangle$ denotes thermal averaging. We split the first six nearest neighbors of the cubic lattice into two coordination shells: 4 atoms on sublattice $A$ and 2 atoms on sublattice $B$ [Fig. 1(d)].

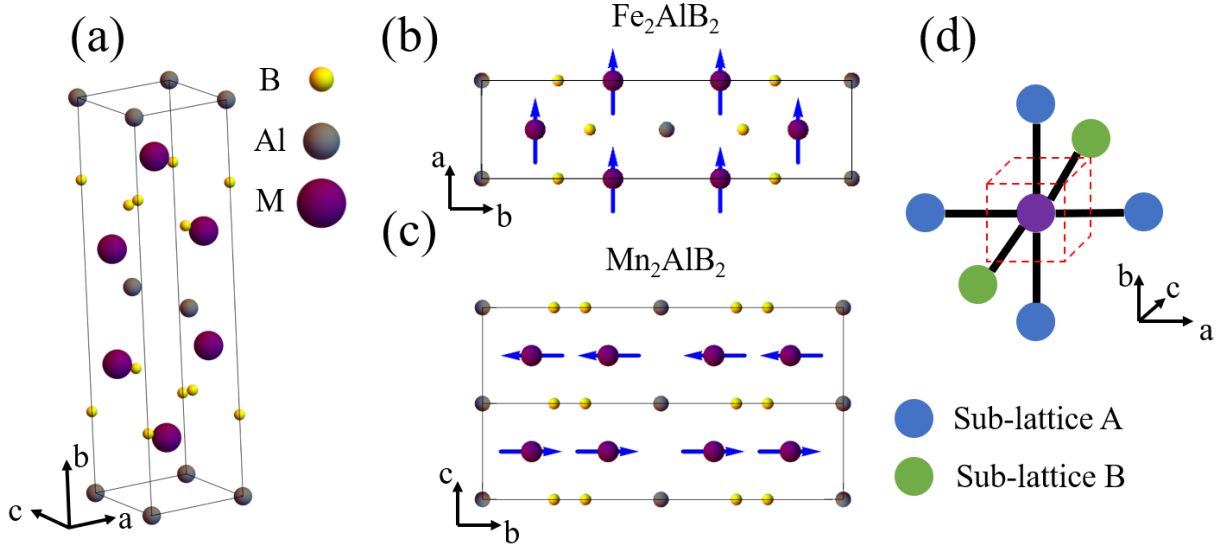

FIG. 1. (a) The chemical unit cell of $M_2\text{AlB}_2$, (b) FM structure of $\text{Fe}_2\text{AlB}_2$,[8] (c) AFM structure of $\text{Mn}_2\text{AlB}_2$,[21] and (d) sublattice structure of simplified mean-field model of $M_2\text{AlB}_2$.

The full Hamiltonian (per atom) of the system is obtained from the single-site Hamiltonians as

$$\hat{H} = \frac{1}{2}(1-x)(\hat{H}_{\text{Fe},A} + \hat{H}_{\text{Fe},B}) + \frac{1}{2}x(\hat{H}_{\text{Mn},A} + \hat{H}_{\text{Mn},B}), \qquad (2)$$

where $\hat{H}_{M,\delta}$ is the single-site Hamiltonian for atom type $M$ on sub-lattice $\delta = A$ or $B$, and is obtained by replacing Fe with Mn and $A$ with $B$ in Eq. (1). To find the magnetization of each atom we need to solve the mean-field self-consistent equations:

$$\langle \hat{S}^\alpha_{M,\delta} \rangle = \text{Tr}\left[\hat{S}^\alpha_{M,\delta} \frac{e^{-(\hat{H}_{M,\delta}/k_B T)}}{Z_{M,\delta}}\right], \quad Z_{M,\delta} = \text{Tr}[e^{-(\hat{H}_{M,\delta}/k_B T)}], \quad \begin{matrix} M = \text{Fe, Mn} \\ \delta = A, B \end{matrix}, \qquad (3)$$

where $T$ is the sample temperature and $k_B$ is Boltzmann's constant. Equation (3) can be expressed as

$$\langle \hat{S}^\alpha_{M,\delta} \rangle = S_M B_{S_M}\left(\frac{|\boldsymbol{\delta}_M| S_M}{k_B T}\right) \frac{\delta^\alpha_M}{|\boldsymbol{\delta}_M|}$$
$$B_S(x) \equiv \frac{2S+1}{2S}\coth\left(\frac{2S+1}{2S}x\right) - \frac{1}{2S}\coth\left(\frac{x}{2S}\right) \qquad (4)$$

where $\boldsymbol{\delta}_M$ is the mean field of atom $M$ on sublattice $\delta$, as defined in Eq. (1). The on-site magnetization is then obtained from

$$M^\alpha(T, x, H^\beta) = \mu_B\left(g^{\alpha\beta}_{\text{Fe}}(1-x)\langle \hat{S}^\beta_{\text{Fe},A} \rangle + g^{\alpha\beta}_{\text{Mn}} x \langle \hat{S}^\beta_{\text{Mn},A} \rangle\right). \qquad (5)$$

The critical temperature is obtained by numerically finding the temperature at which the on-site magnetization vanishes.

## IV. Results and analysis

### A. X-ray powder diffraction

The XRD patterns of $(\text{Fe}_{1-x}\text{Mn}_x)_2\text{AlB}_2$ powders at RT as a function of $x$ are shown in Fig. 2. The reflections are consistent with an orthorhombic phase having the symmetry group *Cmmm* and lattice parameters (LPs) $a \approx 2.9$, $b \approx 11$ and $c \approx 2.9$ Å. Additional reflections belonging to impurity phases are present in small amounts ($\approx 5\%$) for all samples, except $x = 0.5$. In the $x = 0.5$ sample, the $^{11}$B powder was contaminated with $SiO_2$, and thus a relatively large amount of impurities is present. Some of the impurity reflections were identified to belong to $Al_2O_3$ (*R-3c*)[24] and $Fe_4Al_{13}$ (*C2/m*).[25] The XRD patterns were refined using Rietveld refinement as implemented in the FULLPROF package.[26]

Table I. Refined LPs, unit-cell volume ($V$), nominal and refined Mn occupancy ($x$), Debye-Waller factor ($B$) and weight percent of $(Fe_{1-x}Mn_x)_2AlB_2$ phase at room temperature obtained by XRD and NPD. The numbers in parentheses are the standard uncertainties from the Rietveld refinement procedure. A systematic error of 0.03% is estimated between the NPD and XRD LPs and is discussed in Sec. IV C.

| Method | Nominal $x$ | Refined $x$ | a (Å) | b (Å) | c (Å) | V (Å$^3$) | B (Å$^2$) | Wt.% |
|---|---|---|---|---|---|---|---|---|
| XRD | 0[a,c] | N/A | 2.9261(1) | 11.0316(4) | 2.8677(1) | 92.568(5) | --- | 99(1) |
|  | 0.05[a,c] | N/A | 2.9267(1) | 11.0248(4) | 2.8696(1) | 92.589(6) | --- | 98(3) |
|  | 0.1[a,c] | N/A | 2.9281(1) | 11.0269(4) | 2.8774(1) | 92.905(5) | --- | 97(2) |
|  | 0.2[a,c] | N/A | 2.9297(1) | 11.0239(4) | 2.8842(1) | 93.150(6) | --- | 98(2) |
|  | 0.25[a,c] | N/A | 2.92922(4) | 11.0203(2) | 2.88702(4) | 93.195(2) | --- | 97.4(7) |
|  | 0.5[b,c] | N/A | 2.9264(1) | 11.0139(3) | 2.9004(1) | 93.481(5) | --- | 96(2) |
|  | 0.5[b,d] | N/A | 2.92892(8) | 11.0250(3) | 2.90452(7) | 93.791(4) | --- | 87.9(6) |
|  | 0.75[b,c] | N/A | 2.9263(2) | 11.0372(5) | 2.8998(2) | 93.657(8) | --- | 81(2) |
|  | 0.75[b,d] | N/A | 2.92733(5) | 11.0446(2) | 2.90188(5) | 93.821(3) | --- | 85.1(4) |
|  | 1[a,c] | N/A | 2.92025(7) | 11.0613(3) | 2.89568(7) | 93.536(4) | --- | 97(2) |
|  | 1[a,21] | N/A | 2.92267(3) | 11.0715(1) | 2.89776(3) | 93.767(2) | --- | 68.3(5) |
| NPD | 0[a,e] | 0 | 2.92526(2) | 11.0330(1) | 2.86767(3) | 92.552(1) | 0.06(4) | 99(1) |
|  | 0.1[a,e] | 0.096(4) | 2.92746(3) | 11.0287(1) | 2.87765(3) | 92.908(2) | 0.35(3) | 99(1) |
|  | 0.2[a,e] | 0.190(2) | 2.92850(4) | 11.0237(2) | 2.88489(4) | 93.132(2) | 0.42(3) | 99(1) |
|  | 0.25[a,f] | 0.228(4) | 2.9290(3) | 11.019(1) | 2.8876(3) | 93.19(2) | 0.44(4) | 98(1) |
|  | 0.5[a,f] | 0.461(3) | 2.9328(2) | 11.029(1) | 2.9062(2) | 94.00(1) | 0.44[g] | 59(1) |
|  | 1[a,21] | 1 | 2.9166(6) | 11.048(3) | 2.8930(6) | 93.22(4) | 1.1(1) | 100 |

[a] $^{11}B$ sample
[b] Natural boron sample
[c] Measured using a Rigaku X-ray diffractometer
[d] Measured using a Bruker X-ray diffractometer
[e] Measured using the BT-1 neutron diffractometer
[f] Measured using the KANDI-II neutron diffractometer
[g] Fixed using the value for $x = 0.25$ to avoid divergence.

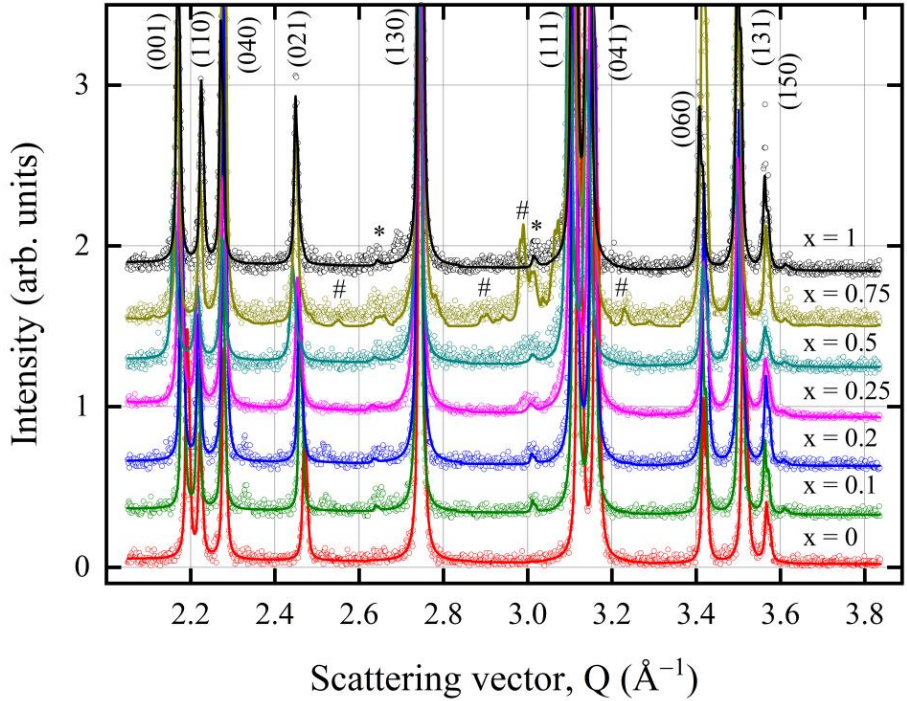

FIG. 2. Observed XRD patterns (symbols) and the corresponding Rietveld refinement (solid lines) for different $(Fe_{1-x}Mn_x)_2AlB_2$ powders with various $x$ values. Reflections are labeled by their Miller indices; impurity reflections are marked by * for $\alpha$-$Al_2O_3$ and # for $(Fe_{1-y}Mn_y)_4Al_{13}$. The patterns for $x = 0.5$ and 0.75 were measured on a natural B sample. All patterns shown here were obtained using a Rigaku diffractometer.

The refined profile consisted of the main orthorhombic phase as well as $\alpha$-$Al_2O_3$ for all samples. The $(Fe_{1-y}Mn_y)_4Al_{13}$ reflections were found in the $x = 0.5$ and 0.75 samples with natural B (Figures S1 and S2 in the SM) and added to the refined profile. The refined parameters for the main phase were the LPs (Table I) and the atomic $y$ positions at the $4j$ and $4i$ sites. The overall Debye-Waller factor could not be refined due to the limited $Q$ range of the diffractometer. Additional reflections present at $Q \approx 2.33$ and $2.52$ Å$^{-1}$ (for $x = 0.1$ and 0.2) and $Q \approx 2.7$ Å$^{-1}$ (for $x = 1$) were not found to belong to any phase containing Fe, Mn, Al, B or any of their oxides.

The obtained LPs (Table I, Fig. 3) vary nonlinearly and nonmonotonically with $x$, and deviate considerably from Vegard's law.[27] The unit-cell volume expands from $\approx 92.5$ Å$^3$ for $Fe_2AlB_2$ up to $\approx 93.5$ Å$^3$ for $Mn_2AlB_2$. These results agree with previous reports by Cedervall *et al.*[13] but are lower than those reported by Chai *et al.*[11] The $a$ and $c$ LPs expand before contracting, while the $b$ LP contracts before expanding. The transition point in all cases is for $x$ in the range 0.2–0.5. The deviation of the LPs from Vegard's law for large $x$ is attributed to magnetostriction within the sample, since for $x \gtrsim 0.5$ the sample becomes AFM at RT, as will be shown in Sec. IV C.

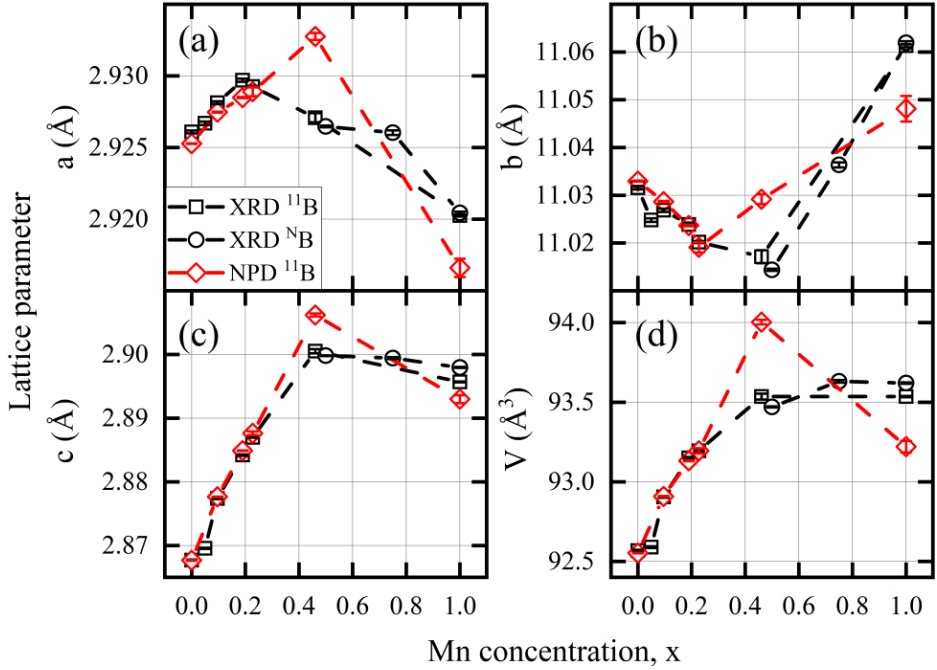

FIG. 3. Refined LPs and unit-cell volumes of $(Fe_{1-x}Mn_x)_2AlB_2$ powders at RT as function of $x$ obtained from XRD (black symbols) and NPD (red symbols). (a), (b), and (c) show the $a$, $b$, and $c$ LPs, respectively and (d) unit-cell volume. Circles indicate measurements performed on natural B samples; squares indicate those performed on $^{11}$B samples. Samples measured with NPD are plotted using the refined $x$. The discrepancies between the NPD and XRD measurements of the same samples originate from systematic errors which are discussed in Sec. IV C.

## B. Magnetization measurements

The temperature dependent magnetization curves (Fig. 4) show varying magnetic responses for different $x$ values. As temperature decreases, samples with $x < 0.5$ show an abrupt increase in magnetization, as expected for a FM. For $x = 0.5$ the increase in magnetization is not as abrupt, while for $x = 0.75$ and $x = 1$ the total magnetic moment is two orders of magnitude lower.

Extrema in the derivative of the magnetization (Fig. 5) are used to determine temperatures of possible magnetic events and the large minima for samples with $x < 0.5$ are used to estimate the critical temperature ($T_C$) for the FM phase. As $x$ increases, additional extrema appear in the derivative [Fig. 5(a), inset]. The origin of these extrema is unclear. However, since the samples with high Mn content also contained more impurity phases, it is possible that these extrema are due to the latter.

The saturated average magnetic moment at 2 K (Table II) is obtained from the high field magnetization [Fig. 6(a)] by linear extrapolation of $M$ as function of $1/H$ curves to $H = 0$ (not shown). The number of data points to include in the linear fit was reduced until the sum of squared residuals ($\chi^2$) did not change. The field-dependent measurement performed at temperatures below and above $T_C$ are used to give a better estimate of $T_C$ by using an Arrott plot (Fig. S3 in the SM) as described in Ref. 28.[18] The new estimates for $T_C$ (Table II) show a systematic increase of ≈4% compared to estimates obtained from d$M$/d$T$ (not shown).

To estimate the magnetocaloric properties of the sample, the isothermal entropy change [Table II, Fig. 6(b)] is calculated by numerically integrating the Maxwell relation:

$$\Delta S_m(T,H) = \int_0^H \left(\frac{\partial M}{\partial T}\right)_{H'} dH' \approx \sum_{i=0}^{n-1} \frac{M_i - M_{i-1}}{T_i - T_{i-1}}(H_i - H_{i-1}) \tag{6}$$

The maximum relative cooling power (RCP, Table II) is estimated by multiplying the maximal value of $\Delta S_m$ by the full width at half maximum (FWHM) of the measured $\Delta S_m$ curve as a function of T [Fig. 6(b)].[29] The calculated RCP of $Fe_2AlB_2$ for a field change of 0–2 T and 0–5 T are 75 and 210 J/kg, respectively. These values are in agreement with results obtained in the literature.[30,31]

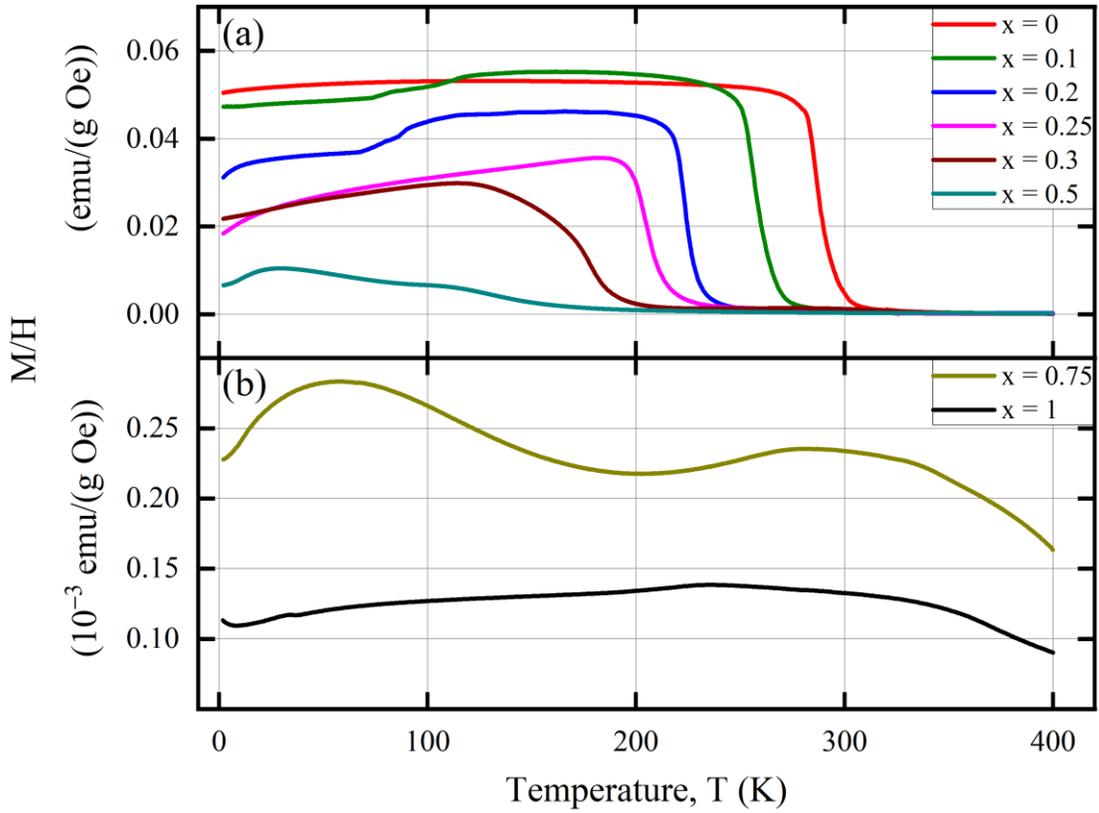

FIG. 4. ZFC magnetization for (a) $x \leq 0.5$, and (b) $x > 0.5$. Measurements for $x = 0.3$, 0.5, and 0.75 were performed on natural B samples. Error bars are smaller than line widths. Note: emu/(g Oe) = $4\pi \times 10^{-3}$ m$^3$/kg.

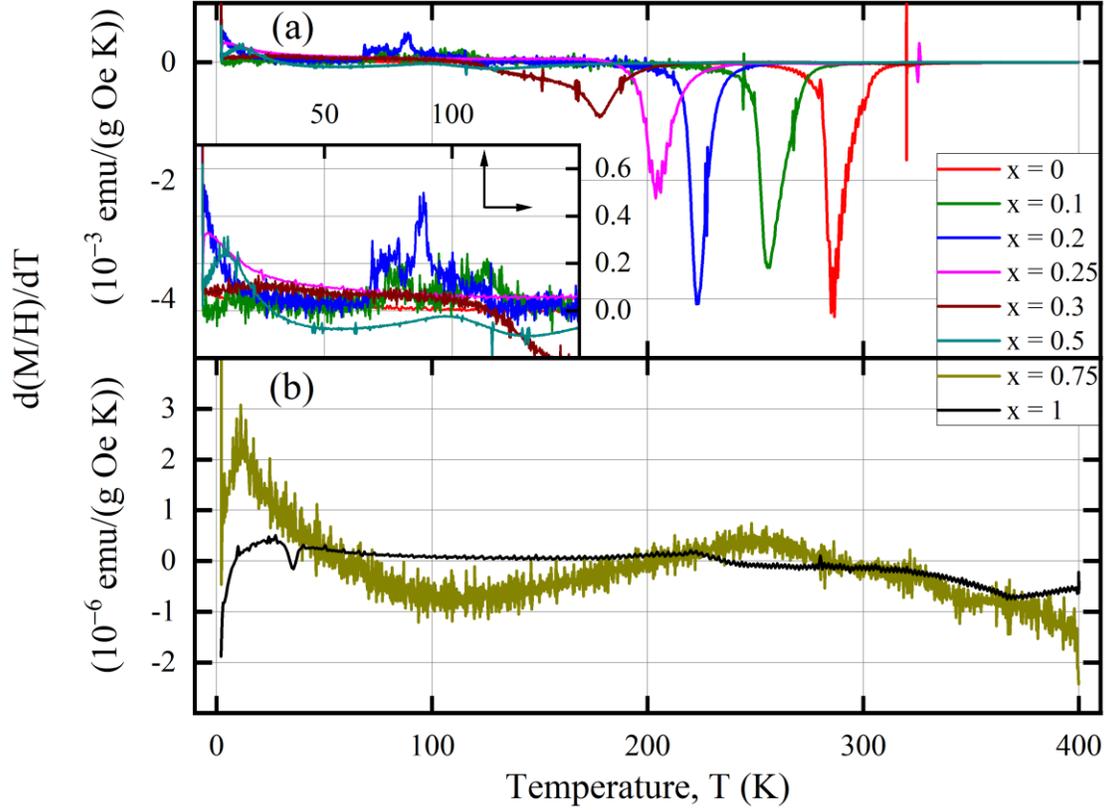

FIG. 5. Derivatives of the ZFC magnetization for (a) $x \leq 0.5$ and (b) $x > 0.5$. Inset in (a) shows a zoomed-in view of the low-temperature regions. Measurements for $x = 0.3, 0.5$, and $0.75$ were performed on natural B samples. Note: emu/(g Oe) = $4\pi \times 10^{-3}$ m$^3$/kg.

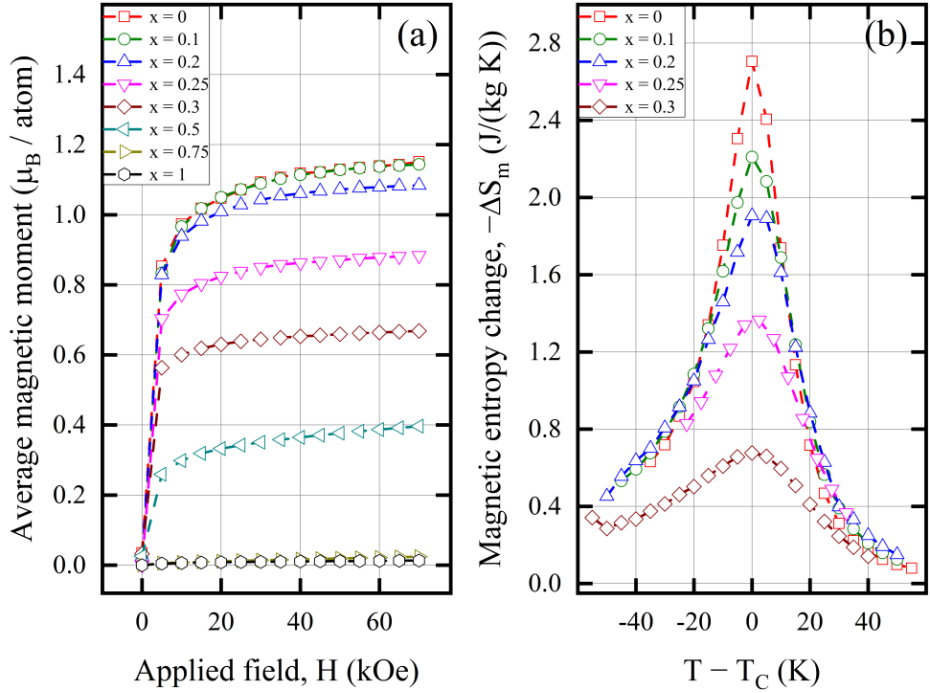

FIG. 6. (a) Field-dependent average magnetic moment of $(Fe_{1-x}Mn_x)_2AlB_2$ at 2 K as function of $x$. (b) Isothermal magnetic entropy change for a field change of 0–20 kOe as function of the relative temperature. Measurements for $x$ = 0.3, 0.5, and 0.75 were performed on natural B samples. The standard errors are smaller than the symbol size. Note: $\mu_0$ Oe = $10^{-4}$ Tesla.

### C. Neutron powder diffraction

The majority of observed reflections in the NPD of all samples at the respective highest measured temperature [cf. Fig. 7(a) for $Fe_2AlB_2$] are consistent with an orthorhombic (*Cmmm*) phase having LPs $a \approx 2.9$ Å, $b \approx 11$ Å, and $c \approx 2.9$ Å. Rietveld refinement of the observed NPD patterns from this sample consisted of a single phase having an orthorhombic (*Cmmm*) symmetry with the starting LPs mentioned above. The atomic $y$ positions of the 4$j$ and 4$i$ sites, as well as the overall Debye-Waller factor, were also refined. Instrumental resolution parameters, zero shift of the detector angle (2$\theta$) and the Mn occupancy in the 4$j$ site were refined for the NPD data at the highest measured temperature and then fixed for all subsequent refinements.

A similar analysis was performed for the other samples measured on BT-1. The instrumental resolution of the KANDI-II diffractometer was determined using a Si standard and was fixed for the refinement of the $x$ = 0.25 and 0.5 samples. The refinement then consisted of the same steps as for the BT-1 samples. In general, the refined RT LPs (Table I) are in agreement with the LPs obtained from XRD, however some deviation, which is larger than the reported statistical uncertainty, is observed. The NPD $a$ LP is lower than that of the XRD LP by ≈0.03% for $x$ = 0, 0.1, and 0.2 while the $b$ and $c$ LPs are overestimated by ≈0.02%.

Table II. Transition temperatures ($T_C$ and $T_N$), saturated average magnetic moment at 2 K ($M_{sat}$), magnetic entropy change ($\Delta S_m$), and relative cooling power (RCP). Ordered FM ($\mu_{FM}$) and AFM ($\mu_{AFM}$) moments of (Fe$_{1-x}$Mn$_x$)$_2$AlB$_2$ as determined by NPD at base temperature. Numbers in brackets indicate uncertainty. The systematic error in $\Delta S_m$ and RCP is expected to be on the order of 10%.[28]

| x | $T_C$ (K)[a] | $T_N$ (K)[b] | $M_{sat}$ ($\mu_B$) | -$\Delta S_m$ (J/kg K) 2 T/5 T | RCP (J/kg), 2 T/5 T | $\mu_{FM}$ ($\mu_B$) | $\mu_{AFM}$ ($\mu_B$) |
|---|---|---|---|---|---|---|---|
| 0[c] | 292.4(2) | --- | 1.19(6) | 2.7/5.7 | 75/210 | 1.30(4) | 0 |
| 0.096(4)[c] | 264.6(3) | --- | 1.18(6) | 2.2/4.6 | 79/218 | 1.25(5) | 0 |
| 0.190(2)[c] | 231.4(3) | 80 ± 20 | 1.12(6) | 1.9/4.0 | 80/226 | 1.07(4) | 0.24(4) |
| 0.228(4)[c] | 212.25(5) | 150 ± 100 | 0.9(1) | 1.4/2.9 | 70/190 | 0.97(6) | 0.54(2) |
| 0.30(2)[d] | 183.16(6) | --- | 0.70(4) | 0.7/1.4 | 41/117 | --- | --- |
| 0.461(3)[c] | --- | 350 ± 50 | --- | --- | --- | 0 | 0.83(2) |
| 0.50(5)[d] | 130(5) | --- | 0.454(3) | --- | --- | --- | --- |
| 1[21,c] | --- | 313[9] | --- | --- | --- | 0 | 0.71(2) |

[a] Critical temperature of the FM component as determined from Arrott plots.
[b] Critical temperature of the AFM component as estimated from NPD measurements. The uncertainties marked with ± indicate upper and lower bounds.
[c] $^{11}$B sample
[d] Natural B sample

Since these discrepancies show the same trend for three different samples, it is safe to assume that they originate from a calibration discrepancy between the XRD and NPD diffractometers. The deviation in the LPs at $x = 0.5$ is attributed to the poor sample quality, which affects the refinement of the LPs due to the low resolution of the KANDI-II diffractometer, and the deviation for $x = 1$ was found to originate from a systematic error in the calibration of the E6 diffractometer.[21] In general, the Mn occupancy is in good agreement with the nominal compositions of the samples, which shows that the (Fe$_{1-x}$Mn$_x$)$_2$AlB$_2$ system is thermodynamically stable over the whole Mn concentration range.

For Fe$_2$AlB$_2$, below 310 K, an increase in the intensity of the (001) reflection [Fig. 7(a), inset] is observed, and is consistent with the onset of FM order, which was shown earlier (see Sec. IV B). We therefore performed an additional refinement of the NPD data (for all temperatures) which included a magnetic phase, with the Fe spins aligned along the crystallographic $a$ axis. Above 290 K, the fit agreement factor for the magnetic phase ($R_{mag}$) shows a large decrease in fit quality, while a refined moment of 0.3 $\mu_B$ is obtained. We therefore take this value as the sensitivity limit for a FM moment of the (Fe$_{1-x}$Mn$_x$)$_2$AlB$_2$ system in the BT-1 diffractometer.

A similar analysis was performed for other compositions. A FM phase was added to all refinements at $T < 260$ K for $x = 0.1$ (Fig. S4 in the SM) and at $T < 220$ K for $x = 0.2$. Below 100 K, an additional reflection appeared at $Q \approx 1.08$ Å$^{-1}$ [Fig. 7(b)] in the NPD data of the (Fe$_{0.8}$Mn$_{0.2}$)$_2$AlB$_2$ sample. This reflection was identified to be the same AFM configuration found in $Mn_2AlB_2$ (see Sec. II). The refinement therefore contained an AFM phase for all measurements of $(Fe_{0.8}Mn_{0.2})_2AlB_2$ with $T < 100$ K. The sensitivity limit of BT-1 for an AFM moment was determined to be 0.2 $\mu_B$ in the same manner as for the FM moment. For the $(Fe_{0.75}Mn_{0.25})_2AlB_2$ composition, the refinement at 3 K contained both an FM and AFM phases (Fig. S5 in the SM).[18] The sensitivity limits of the KANDI-II diffractometer were determined by refining both magnetic phases at RT for the $x = 0.25$ sample, and were found to be 0.4 $\mu_B$ and 0.2 $\mu_B$ for the FM and AFM moments, respectively. The diffraction pattern of $(Fe_{0.5}Mn_{0.5})_2AlB_2$ at RT showed an excess neutron count at the position of the (0, 0, 1/2) reflection (Fig. S6 in the SM),[18] which did not appear in the XRD pattern, excluding the possibility for an impurity phase. An attempt to add a FM phase to the refinement did not change the values for the refined parameters, showing no correlation between the FM moment and other parameters. The FM moment of 0.45 $\mu_B$, observed by magnetization measurements (Table II) could not be detected by NPD for this sample. Therefore, although both the FM and AFM phases are present, only the AFM phase was included in the refinement for this sample.

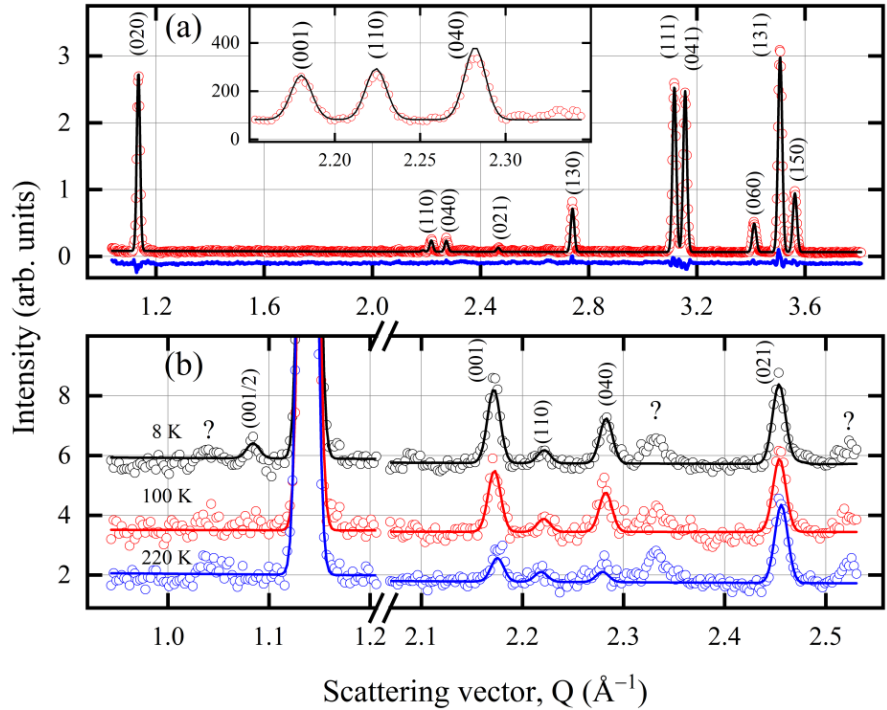

FIG. 7. (a) Observed NPD pattern of $Fe_2AlB_2$ powders (symbols) at 350 K, the corresponding Rietveld refinement (solid line) and their difference (blue bottom solid line). Inset zooms in on the FM reflection at 8 K. (b) Observed NPD (symbols) of $(Fe_{0.8}Mn_{0.2})_2AlB_2$ at different temperatures and the corresponding Rietveld refinements (solid line). Reflections are marked using their Miller indices and fractional Miller indices (for AFM reflections). Reflections marked by "?" correspond to unidentified impurity phases. The measurements were performed using the BT-1 diffractometer. The error bars are smaller than the symbol size.

The temperature evolution of the LPs for $Fe_2AlB_2$ [Fig. 8(a)] shows an expansion of the $c$ LP upon cooling below 310 K. Combined with the onset of FM ordering below this temperature, it is reasonable to conclude that this anomalous thermal expansion most likely originates from magnetostriction. These results agree with density-function theory (DFT) calculations by Ke *et al.*,[32] that have shown a strong dependence of the magnetic moment in $Fe_2AlB_2$ on the $c$ LP. The changes in the LPs over most of the $x$ range are of the order of 0.25%. A similar behavior is observed for the $x = 0.1$ and 0.2 samples (Fig. S7 in the SM),[18] while an expansion of the $b$ LP upon cooling is observed for $x = 1$.[21] For $x = 0.5$ [Fig. 8(b)], the $c$ LP contracts below 200 K and expands below 100 K. This change in behavior is attributed to the FM transition observed at ≈130 K (Table II).

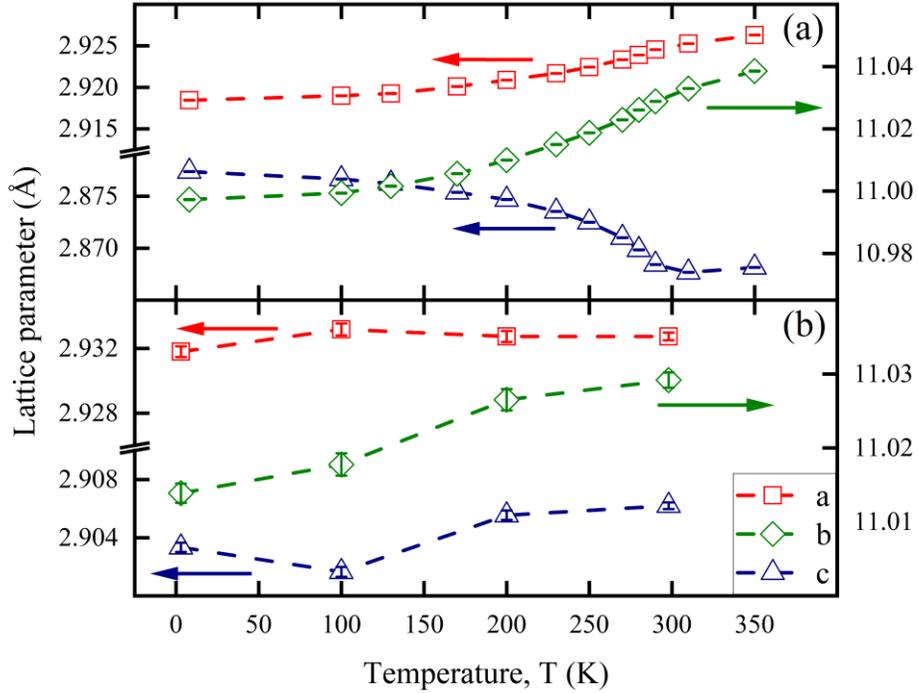

FIG. 8. Temperature evolution of the $a$ and $c$ LPs (left $y$ axis) and $b$ (right $y$ axis) of (a) $Fe_2AlB_2$ and (b) $(Fe_{0.5}Mn_{0.5})_2AlB_2$ samples determined from NPD patterns.

## V. Mean field theory analysis

The observed magnetic reflections (Fig. 7) indicate that the magnetic structure of $(Fe_{1-x}Mn_x)_2AlB_2$ is composed of two parts: a FM moment, which points along the crystallographic $a$ axis, and an AFM moment, which points along the crystallographic $b$ axis with a propagation vector of $\mathbf{k} = (0, 0, 1/2)$. The temperature evolution of each magnetic component [Fig. 9(a), symbols] shows a gradual decrease, typical of a second-order phase transition. The FM and AFM components, when present, have different critical temperatures and ground-state magnitudes, that vary with $x$ (Table II). The phase diagram of $(Fe_{1-x}Mn_x)_2AlB_2$ [Fig. 9(b), symbols] thus consists of three different phases. For $x < 0.23$, only a FM phase is present. For $0.23 < x \leq 0.46$, both FM and AFM phases are present, while for $0.46 < x \leq 1$ only an AFM phase is present.

To investigate the magnetic moment dependence on $x$ and $T$, we made use of Eq. (5). The unknown parameters in the model are the $g$ factors and spins of the Fe and Mn atoms, and the exchange constants. Since the FM component is directed along the $a$ axis, while the AFM component is directed along the $b$ axis, we only need to consider the exchange constants along these directions. This leaves us with four exchange constants, namely: $J_{\text{Fe-Fe}}^{(c),xx}$, $J_{\text{Fe-Mn}}^{(c),xx}$, $J_{\text{Fe-Mn}}^{(c),yy}$, and $J_{\text{Mn-Mn}}^{(c),yy}$. We assume the $g$ factors of the two atoms to be isotropic, i.e., $g_I^{\alpha\beta}=g_I^0\delta^{\alpha\beta}$. The fitting procedure is obtained as follows. The values of $S_{\text{Fe}}$ and $S_{\text{Mn}}$ are scanned in the range 0.5–3 in steps of 1/2. For each pair ($S_{\text{Fe}}$, $S_{\text{Mn}}$) $g_{\text{Fe}}$ and $J_{\text{Fe-Fe}}^{(c),xx}$ are obtained by fitting the temperature evolution of the ordered magnetic moment, $\mu(T)$ for Fe$_2$AlB$_2$; $g_{\text{Mn}}$ and $J_{\text{Mn-Mn}}^{(c),yy}$ are obtained by fitting $\mu(T)$ for Mn$_2$AlB$_2$ [Fig. 9(a)]. Next, $J_{\text{Fe-Mn}}^{(c),xx}$ and $J_{\text{Fe-Mn}}^{(c),yy}$ are fitted to best match $\mu(T)$ for $x$ = 0.1, 0.2, 0.25, and 0.5 [Fig. 9(a), solid line]. The $\chi^2$ goodness of fit parameter is used to identify the best-matching fit, while also requiring that the resulting values for the exchange parameters remain positive. The entire fitting procedure was performed twice where $J_{\text{Fe-Mn}}^{(c),xx}$ and $J_{\text{Fe-Mn}}^{(c),yy}$ were assumed to be FM or AFM along the $c$ axis. Finally, the best-matching parameters were obtained by calculating $\mu(x)$ at base temperature (Fig. 10). The only parameter set which predicted the existence of a nonzero FM moment for $x$ = 0.5 was $S_{\text{Fe}}$ = 3/2, $S_{\text{Mn}}$ = 1/2, $g_{\text{Fe}}$ = 0.86(2), and $g_{\text{Mn}}$ = 1.38(1). The values for the exchange constants (in meV) are $J_{\text{Fe-Fe}}^{(c),xx}$ = 3.67(8), $J_{\text{Fe-Mn}}^{(c),xx}$ = 2.3(5), $J_{\text{Fe-Mn}}^{(c),yy}$ = 11.7(4), and $J_{\text{Mn-Mn}}^{(c),yy}$ = 18.9(2). The sign of $J_{\text{Fe-Fe}}^{(c),xx}$ and $J_{\text{Fe-Mn}}^{(c),xx}$ is positive along all directions, while the sign of $J_{\text{Fe-Mn}}^{(c),yy}$ and $J_{\text{Mn-Mn}}^{(c),yy}$ is negative along the $c$ axis.

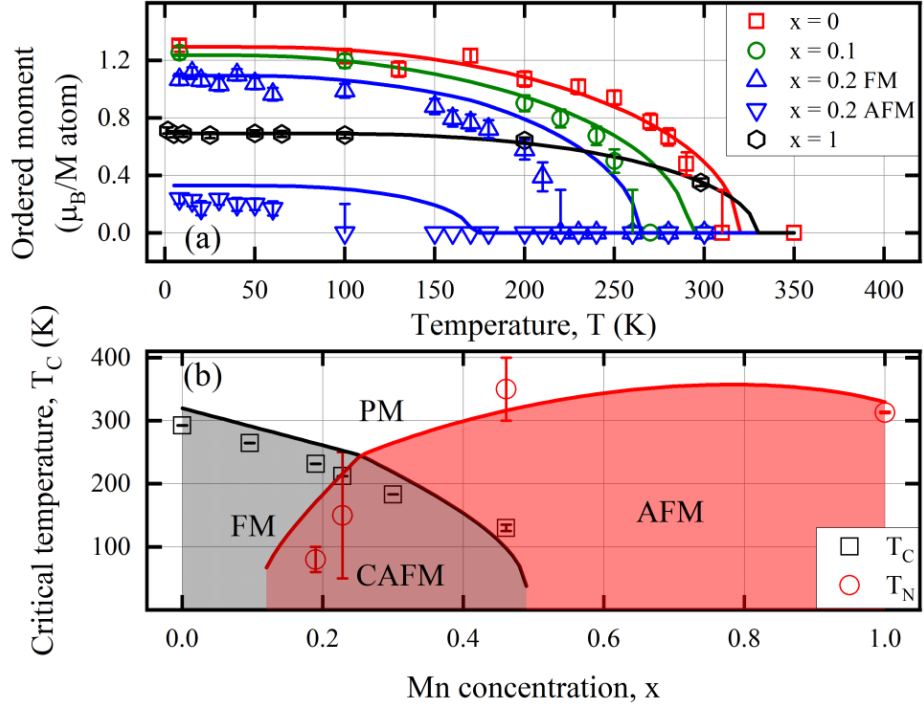

FIG. 9. (a) Temperature evolution of observed total ordered magnetic moment (symbols) in $(Fe_{1-x}Mn_x)_2AlB_2$. (b) Observed (symbols) and calculated (solid lines) critical temperature of FM (black) and AFM (red) components as function of $x$. Different regions in the phase diagram are labeled by magnetic phases present in them.

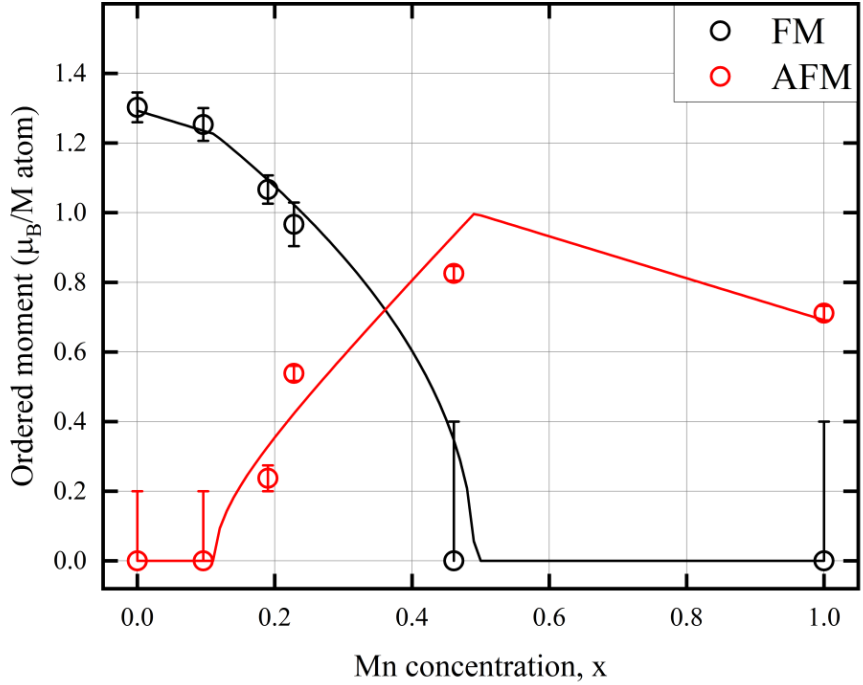

FIG. 10. Observed ordered magnetic moments at base temperature as function of *x*. Solid lines are fits of Eq. (5).

## VI. Discussion

The calculated magnetic phase diagram of the solid solution $(Fe_{1-x}Mn_x)_2AlB_2$ [Fig. 9(b)] contains three types of ordered magnetic structures: a FM structure below a critical Mn concentration of $x_1 \approx 0.1$, an AFM structure above $x_2 \approx 0.5$, and a combination of both in between. This intermediate region is interpreted as a canted AFM. Because the LPs of $Fe_2AlB_2$ and $Mn_2AlB_2$ differ significantly (Table I), a separation of the sample into Fe-rich and Mn-rich clusters would produce two distinctly visible diffraction patterns. Since only a single diffraction pattern is observed, with no broadening of the crystallographic or magnetic reflections relative to the instrumental resolution, we conclude that the mixing of Mn in the sample is homogeneous and that the observed combination of FM and AFM structures is to be interpreted as a canting of the FM moments. The canting angles, in the *a-b* plane relative to the *b*-axis, at base temperature are estimated to be 13(2) ° and 29(2) ° for $x = 0.19$ and 0.23, respectively. The general features of this phase diagram are qualitatively well described by MFT [Eq. (5)], although quantitative agreement is far from perfect. Previous DFT calculations have concluded that the AFM configuration becomes more stable than the FM configurations for $x > 0.2$.[32] This result agrees with the observed NPD results, but places a higher bound on the critical *x* than MFT. We note that unlike previous reports,[12,13] no evidence for a disordered magnetic phase was found.

The overestimation of $T_C$ and $T_N$ in the calculated model, may partly be a result of the mean-field approximation, which is known for giving overestimates for critical temperatures.[33] The best-fitted absolute values for the exchange constants are similar to values that were computed by DFT;[13,32] however direct comparison is difficult due to the simplifications introduced in the mean-field model. The Fe-Mn and Mn-Mn couplings are found to be negative along the $c$ axis, which is also the shortest axis. This suggests that the magnetic interaction between the Fe and Mn atoms is a direct exchange interaction, since this interaction is known to change sign from FM to AFM with decreasing interatomic distance as described by the Bethe-Slater curve.[34] This suggestion is corroborated by the DFT calculations which have shown that the Mn-Mn exchange coefficients are negative along the $c$ axis but positive along the $a$ axis. Since the latter is longer than the former by only 0.02 Å, we can obtain an estimate on the critical Mn-Mn distance to be in the 2.89–2.92 Å range.

The anomalous variation of the LPs with $T$ (Fig. 8) and $x$ (Fig. 3) indicates a strong magnetoelastic interaction. This variation in interatomic distances in turn influences the strength of the exchange interaction between the magnetic $M$ atoms, giving rise to a complicated dependence of the ordered magnetic moment on $T$ and $x$ (Fig. 9). These subtleties were not considered in our simplified model. In addition, the magnetoelastic interaction in these compounds is highly anisotropic, as can be seen from the qualitatively different temperature evolution of the LPs (Fig. 8). For $x \leq 0.5$ the magnetic moment is highly affected by the $c$ LP, causing an anomalous expansion upon cooling. A similar dependence was observed in $Mn_2AlB_2$ for the $b$ LP and indicated a change in the anisotropy of the magnetoelastic interaction.[21]

The addition of Mn into $Fe_2AlB_2$ decreases the ordered FM moment, which in turn decreases the overall magnetocaloric effect (Table II). However, the maximum in the magnetic entropy change occurs over a broader temperature range [Fig. 6(b)] resulting in a 6% increase in the estimated RCP (Table II). The addition of Mn does not seem to broaden the magnetic transition, as can be observed from the temperature evolution of the ordered magnetic moments [Fig. 9(a)]. Additionally, since, as discussed above, the introduction of Mn does not produce multiple phases in the sample but is admixed homogeneously, we can conclude that the broadening of the MCE curve is not caused by chemical disorder but rather by the introduction of competing AFM interactions, which are theoretically known to broaden the range of the MCE.[17] Addition of 10% Mn decreases $T_C$ from ≈290 K to ≈260 K while the effective temperature range or FWHM of the MCE stays at ≈30 K. This enables control over $T_C$ in the RT range without a substantial loss of cooling power. For example, mixing multiple $(Fe_{1-x}Mn_x)_2AlB_2$ compounds with different $x$ can result in a combined MCE curve with a desired shape, which is controlled by the ratio of different compounds and their respective $T_C$'s.

## VII. Conclusions

The magnetic phase diagram of the quaternary boride, $(Fe_{1-x}Mn_x)_2AlB_2$, was studied using x-ray- and neutron-powder diffraction, and magnetization measurements. In agreement with MFT predictions, this system offers three magnetic ground states at different Mn concentrations: ferromagnetic (FM), antiferromagnetic (AFM), and a canted AFM (Fig. 10).

While the addition of Mn decreases the critical temperature [Fig. 9(b)], FM moment (Fig. 10), and magnetic entropy changes [Fig. 6(b)], it does increase the relative cooling power for Mn additions up to $x \approx 0.2$. This comes about due to the broadening of the temperature range, over which the magnetocaloric effect is significant. It is therefore possible to fine tune the transition temperature of $Fe_2AlB_2$ in the 274–294 K (0–20 °C) range without a considerable loss of cooling power.

# Acknowledgements

D.P. thanks E. Greenberg for help with performing XRD measurements. A.P, A.K, D.P, O.R and E.N.C acknowledge the support of the Israel Atomic Energy Commission Pazy Foundation Grant. H.A.E. thanks the National Research Council (USA) for financial support through the Research Associate Program. S.K., M.S., L.H. and M.W.B. acknowledge the Knut and Alice Wallenberg Foundation (Grant No. KAW 2015.0043).

Certain commercial equipment, instruments, or materials (or suppliers, or software, ...) are identified in this paper to foster understanding. Such identification does not imply recommendation or endorsement by the National Institute of Standards and Technology, nor does it imply that the materials or equipment identified are necessarily the best available for the purpose.

[34] A. Sommerfeld and H. Bethe, *Hanbuch Der Physik* (Springer, Berlin, 1933).

# Supplemental Material

## I. Sample preparation

All samples were prepared using a powder metallurgical route. Elemental Fe (Alfa Aesar, 99.5% metals basis, 6-10 µm, reduced), Mn (Alfa Aesar, 99.6% metals basis, < 10 µm), Al (Alfa Aesar, 99.5% metals basis, < 44 µm), and $^{11}$B powders (Cambridge Isotopes, 99%) were used as reagents. The as-received $^{11}$B was ground in an agate mortar and pestle, sieved to obtain a particle size < 44 µm. The $^{11}$B powders were then washed in a 10 wt.% HF solution for 4 hours at room temperature to remove $SiO_2$ impurites, washed in distilled water, and dried in ambient air before further processing. Some compositions were instead, or additionally, prepared using "natural" B powders (Alfa Aesar, 98.8% metals basis, crystalline, <38 µm) containing a natural abundance of $^{10}$B and $^{11}$B isotopes (denoted as $^N$B for brevity). The methods used to prepare samples of a given $(Fe_{1-x}Mn_x)_2AlB_2$ solid solution composition are summarized in Table SIII. All reaction steps were conducted in an atmosphere of flowing Ar (99.9999% purity) inside a horizontal alumina tube furnace with both Ti and Y powders upstream as oxygen getters.

### A. Method 1

First, Fe, Mn, and B were reacted to form $(Fe_{1-x}Mn_x)B$ ternary boride solid solutions and then reacted with Al to make the quaternary $(Fe_{1-x}Mn_x)_2AlB_2$ solid solutions. 4.5 grams of the elemental powders were mixed in molar ratios of $(1 - x)Fe + xMn + 1.025$ $^{11}$B, wherein $x = (0, 0.05, 0.10$ and $0.20)$. The mixtures were ball milled for 24 hours in plastic jars with an equal mass of yttria-stabilized zirconia media. Green bodies were uniaxially cold pressed at 200 MPa into 2.5 cm wide disks. All formulations were heated at a rate of 5 K/min in a horizontal tube furnace to 1474 K (1200 °C) and held at this temperature for 5 h before cooling at the same rate to room temperature. The reacted $(Fe_{1-x}Mn_x)B$ ternary boride pellets were then crushed with an agate mortar and pestle and sieved to a size < 44 µm. The ternary borides were mixed with Al in molar ratios of $2(Fe_{1-x}Mn_x)B + 1.2Al$ then ball milled and pressed into green bodies as described above. The pellets were heated at a rate of 5 K/min to 1324 K (1050 °C) and maintained at this temperature for 15 h before cooling to room temperature. All samples were then ground and sieved to obtain particles < 44 µm.

### B. Method 2

In this case, elemental powders in the molar ratio $(2 - 2x)Fe + (2x)Mn + 1.2Al + 2$ $^N$B powders were ball milled and cold-pressed into pellets as described above. The pellets were heated at a rate of 5 K/min to 1324 K and maintained at this temperature for 1 h before cooling to room temperature. The loosely sintered pellets were pulverized in an agate mortar and pestle, re-pressed into a pellet, and re-heated again at 1324 K for 15 h. The reacted pellets were pulverized into powders for further characterization and magnetic measurements.

Table SIII. Summary of compositions prepared and the corresponding synthesis method used.

| Nominal x | Boron Reagent | |
| --- | --- | --- |
| | $^{11}B$ | $^{N}B$ |
| 0 | Method 1 | Method 1 |
| 0.05 | Method 1 | N/A |
| 0.1 | Method 1 | N/A |
| 0.2 | Method 1 | N/A |
| 0.25 | Method 2 | Method 2 |
| 0.3 | Method 2 | Method 2 |
| 0.5 | Method 2[a] | Method 2 |
| 0.75 | Method 2 | Method 2 |
| 1 | Method 2[a] | Method 2 |

[a] $^{11}B$ reagent was not purified with HF treatment prior to reaction (i.e. used as-received)

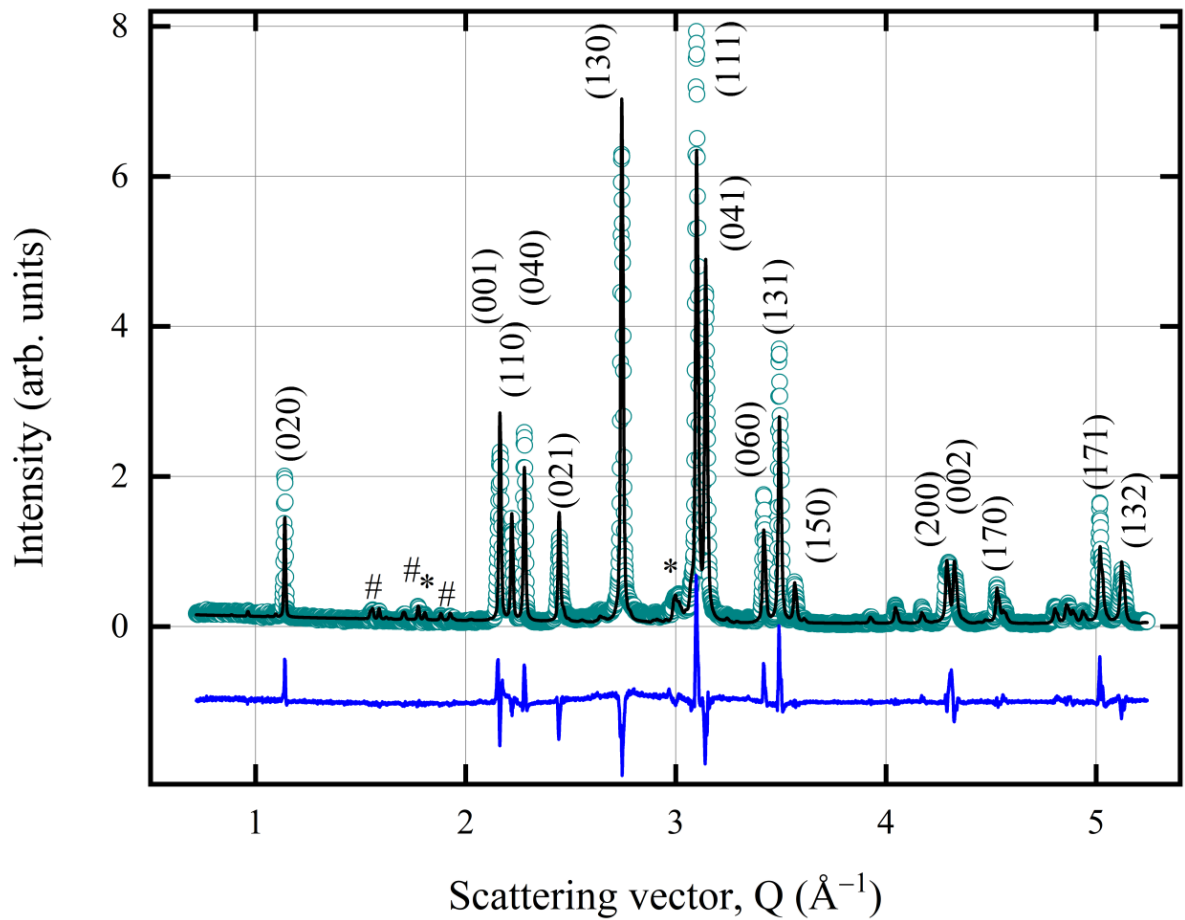

FIG. S1. Observed XRD pattern (symbols) and the corresponding Rietveld refinement (solid line) for a natural B $(Fe_{0.5}Mn_{0.5})_2AlB_2$. Reflections are labeled by their Miller indices; impurity reflections are marked by * for $Al_2O_3$ and # for $(Fe_{1-y}Mn_y)_4Al_{13}$. The measurement was performed using a Bruker D8 – Advance. The error bars are smaller than the symbol size.

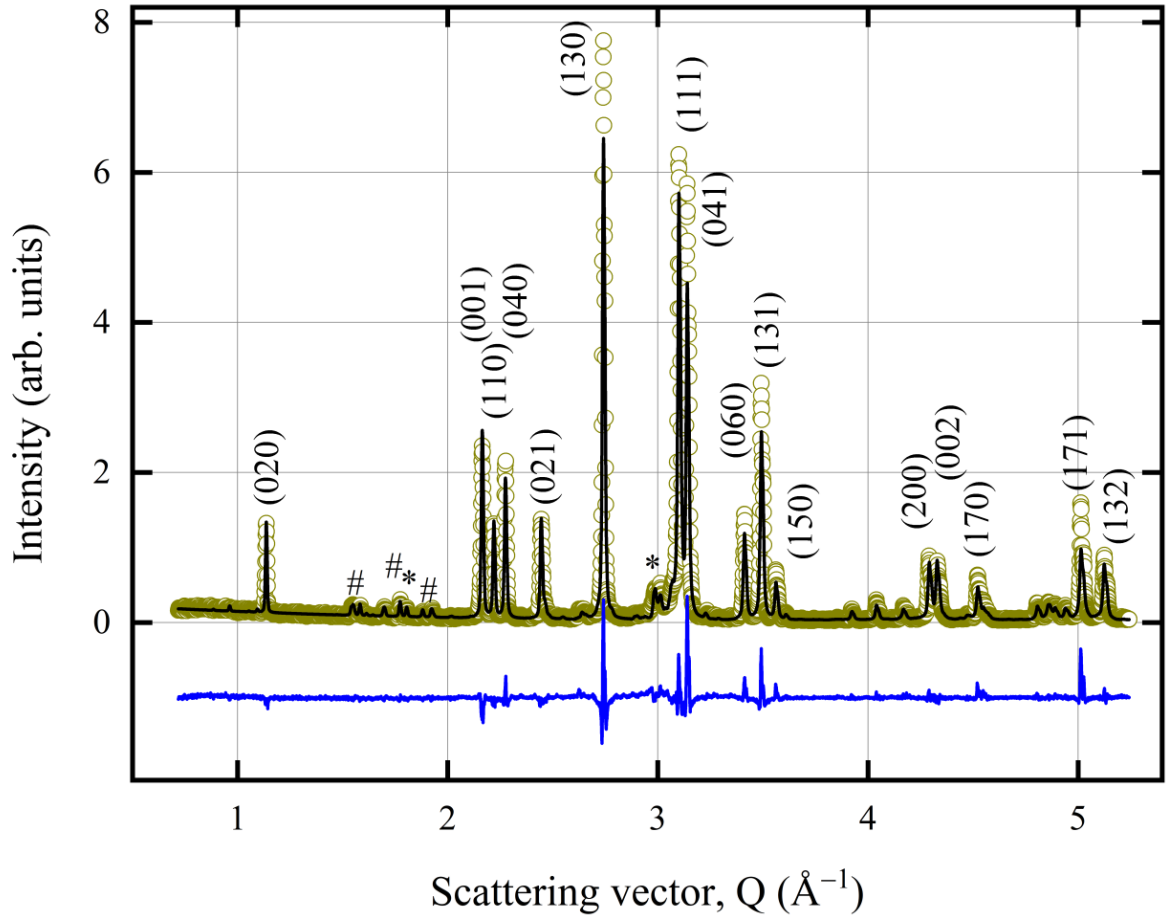

FIG. S2. Observed XRD pattern (symbols) and the corresponding Rietveld refinement (solid line) for a natural B $(Fe_{0.25}Mn_{0.75})_2AlB_2$. Reflections are labeled by their Miller indices; impurity reflections are marked by * for $Al_2O_3$ and # for $(Fe_{1-y}Mn_y)_4Al_{13}$. The measurement was performed using a Bruker D8 – Advance. The error bars are smaller than the symbol size.

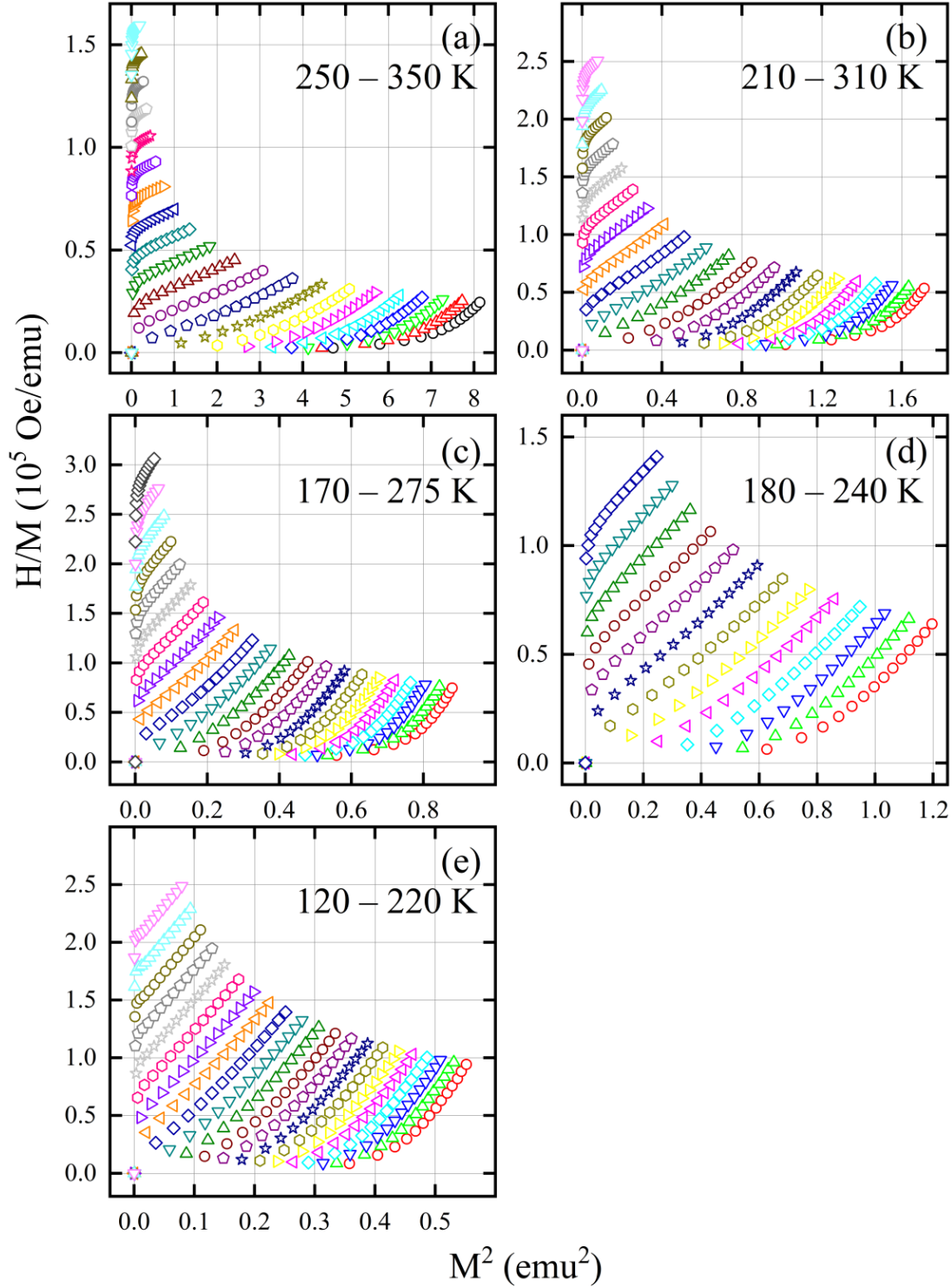

FIG. S3. Arrott plots for $(Fe_{1-x}Mn_x)_2AlB_2$ with $x = 0$ (a), 0.1 (b), 0.2 (c), 0.25 (d), and 0.3 (e). The error bars are smaller than the symbol size. The temperature step in each sample is 5 K. Note: $\mu_0$ Oe = $10^{-4}$ Tesla and emu = $10^{-3}$ A m$^2$.

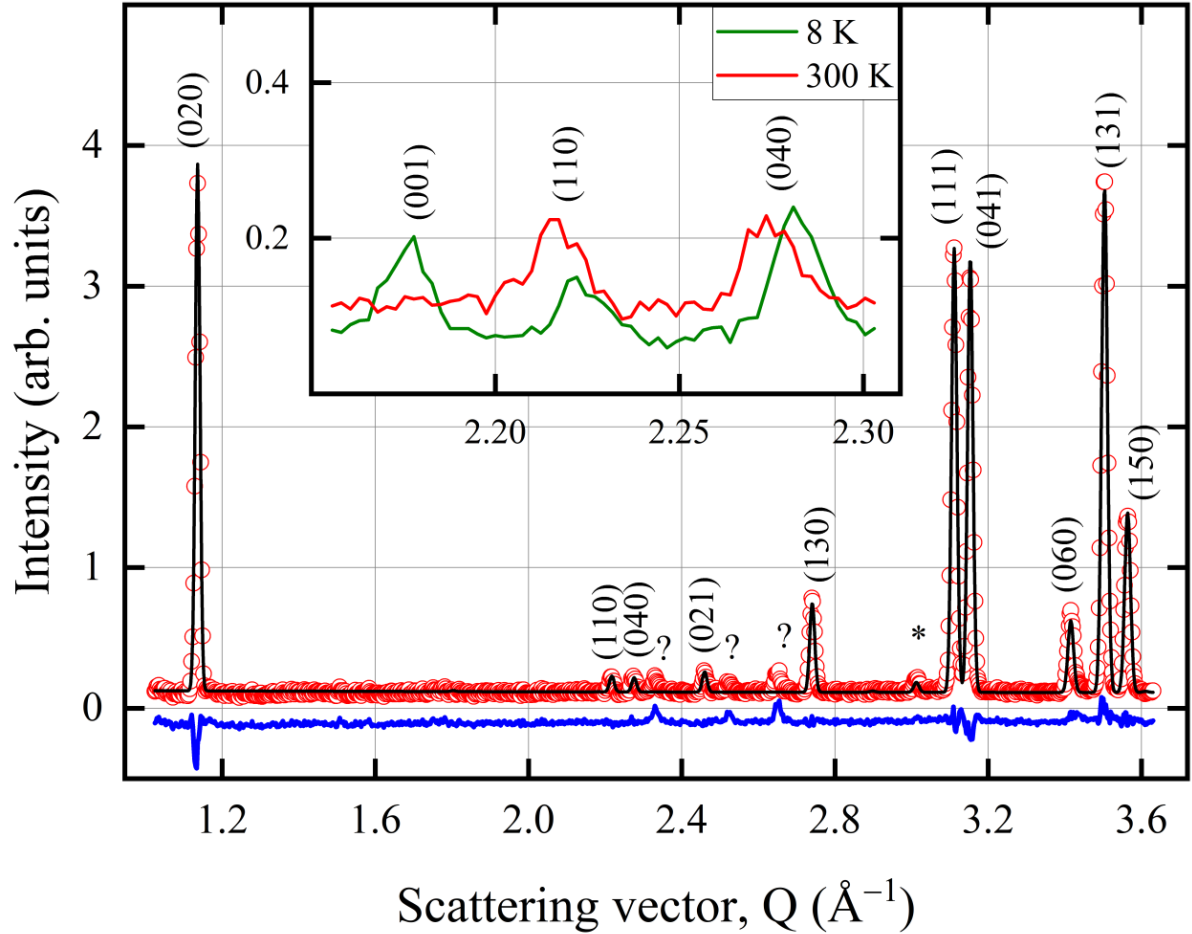

FIG. S4. Observed NPD pattern of $(Fe_{0.9}Mn_{0.1})_2AlB_2$ powders (symbols) at 298 K, the corresponding Rietveld refinement (solid line), and their difference (solid blue line). Reflections are labeled by their Miller indices; impurity reflections are marked by * for $Al_2O_3$ and "?" for unidentified phases. Inset zooms in on the FM (001) reflection that appears below 260 K. The measurement was performed using the BT-1 diffractometer. The error bars are smaller than the symbol size.

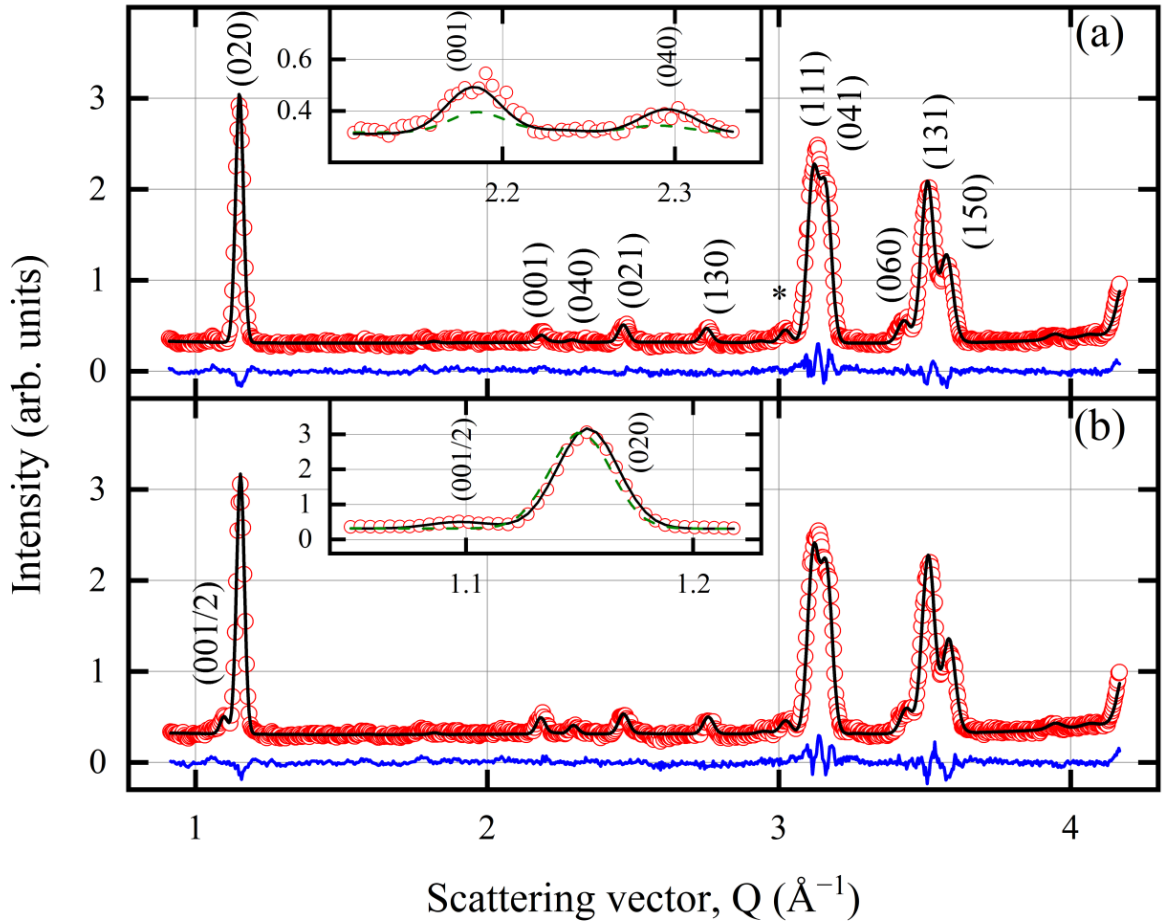

FIG. S5. Observed NPD pattern of $(Fe_{0.75}Mn_{0.25})_2AlB_2$ powders (symbols) at (a) 298 K, (b) 3 K, the corresponding Rietveld refinement (solid line), and their difference (solid blue line). Reflections are labeled by their Miller indices; impurity reflections are marked by * for $Al_2O_3$. Inset in (a) zooms in on the FM (001) reflection, inset in (b) zooms in on the AFM (001/2) reflection. The dashed green line shows the calculated profile at 298 K. The measurement was performed using the KANDI-II diffractometer. The error bars are smaller than the symbol size.

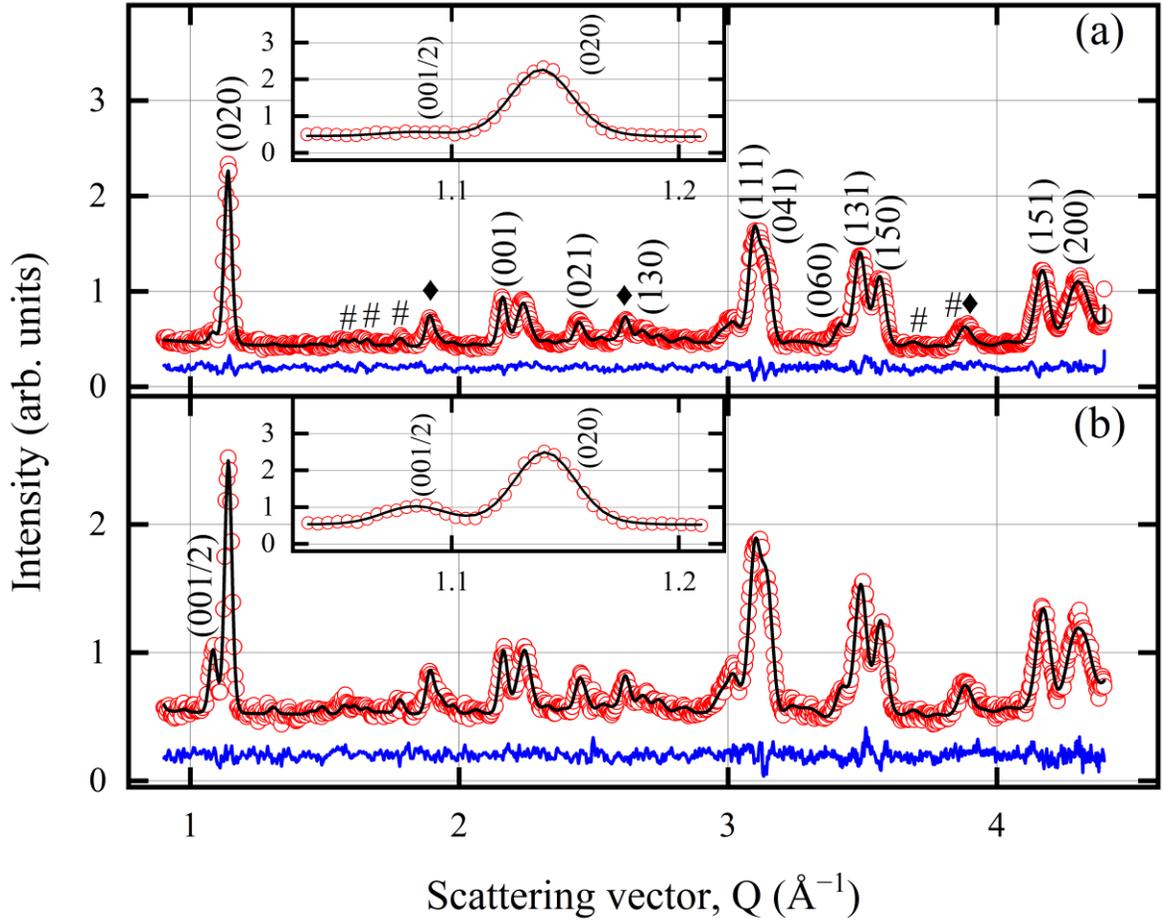

FIG. S6. Observed NPD pattern of $(Fe_{0.5}Mn_{0.5})_2AlB_2$ powders (symbols) at (a) 298 K, (b) 3 K, the corresponding Rietveld refinement (solid line), and their difference (solid blue line). Reflections are labeled by their Miller indices; impurity reflections are marked by # for $(Fe_{1-y}Mn_y)_4Al_{13}$ and ♦ for $(Fe_{1-z}Mn_z)B$. Insets zooms in on the AFM (001/2) reflection. The measurement was performed using the KANDI-II diffractometer. The error bars are smaller than the symbol size.

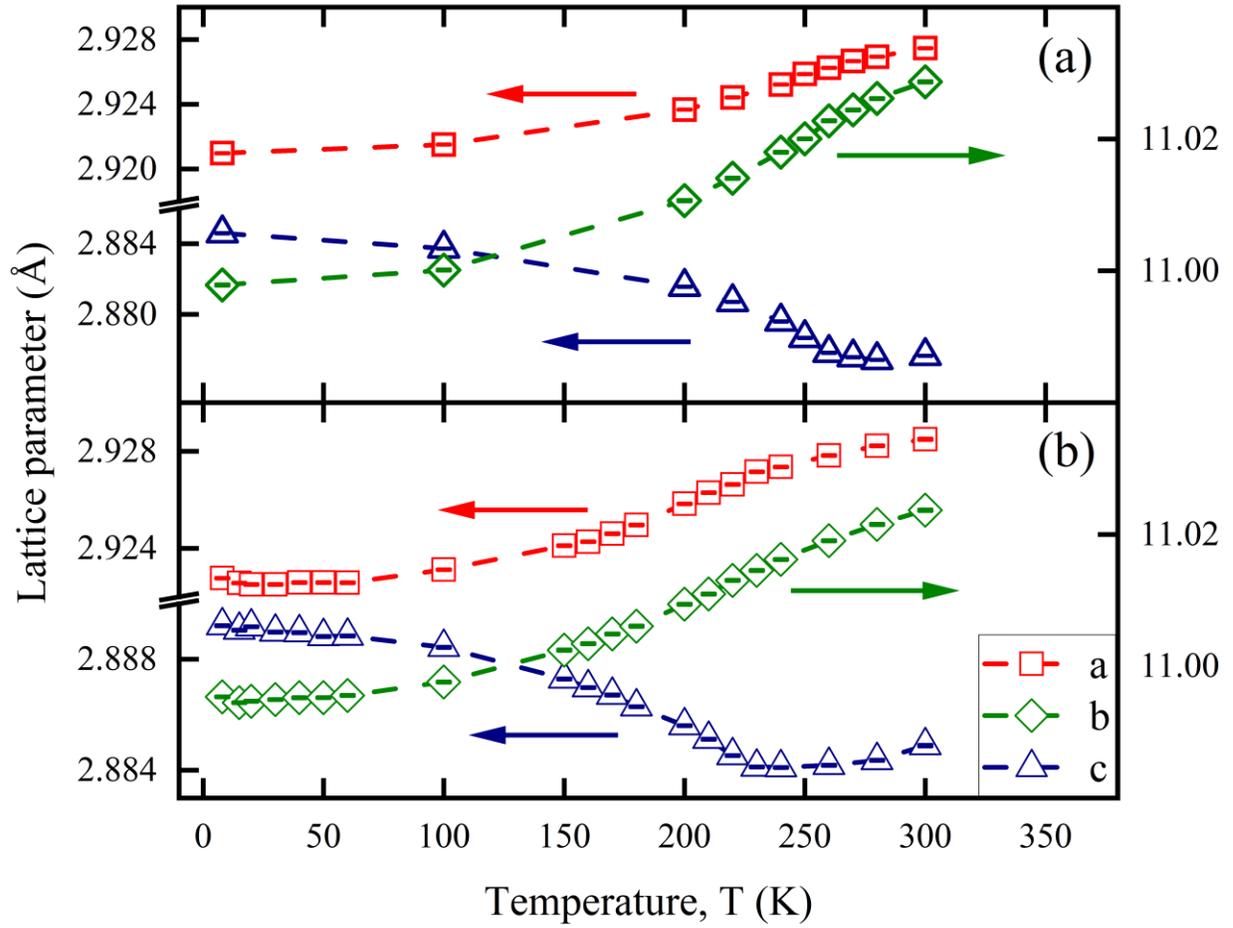

FIG. S7. Temperature evolution of the *a* and *c* LPs (left *y* axis) and *b* (right *y* axis) of (a) $(Fe_{0.9}Mn_{0.1})_2AlB_2$ and (b) $(Fe_{0.8}Mn_{0.2})_2AlB_2$ determined from the NPD patterns.